\documentclass[11pt]{article}

\usepackage[margin=1in]{geometry}
\usepackage{mathpazo}
\usepackage{stmaryrd}
\usepackage[english]{babel}
\usepackage[autostyle, english = american]{csquotes}
\usepackage{anyfontsize}
\usepackage{authblk}

\usepackage{amssymb, amsthm, amsmath, dsfont, mathtools}
\mathtoolsset{showonlyrefs}
\usepackage{latexsym}

\usepackage{eepic}
\usepackage{sectsty}
\usepackage{framed}
\usepackage[shortlabels]{enumitem}
\usepackage{float}
\restylefloat{table}
\usepackage{xifthen,mathrsfs}
\usepackage{ytableau}
\usepackage{subfigure}

\usepackage[pagecolor=white]{pagecolor}
\usepackage{xcolor}
\makeatletter
\newcommand{\globalcolor}[1]{%
  \color{#1}\global\let\default@color\current@color
}
\makeatother

\AtBeginDocument{\globalcolor{black}}

\usepackage{hyperref}
\usepackage{hyperref}
\hypersetup{
    colorlinks = true,
    linkcolor = magenta,
    citecolor = blue,
    urlcolor = blue
}

\hypersetup{pdfpagemode=UseNone}

\newtheorem{theorem}{Theorem}
\numberwithin{theorem}{section}
\newtheorem{lemma}[theorem]{Lemma}
\newtheorem{prop}[theorem]{Proposition}
\newtheorem{cor}[theorem]{Corollary}

\newtheorem*{theorem*}{Theorem}

\theoremstyle{definition}

\newtheorem{remark}[theorem]{Remark}

\newtheorem{example}[theorem]{Example}

\newcommand{\tinyspace}{\mspace{1mu}}

\renewcommand{\t}{{\scriptscriptstyle\mathsf{T}}}

\newcommand{\biggceil}[1]{\biggl\lceil #1 \biggr\rceil}

\newcommand{\norm}[1]{\lVert\tinyspace #1 \tinyspace\rVert}

\newcommand{\ket}[1]{\lvert\tinyspace #1 \tinyspace \rangle}
\newcommand{\bra}[1]{\langle\tinyspace #1 \tinyspace \rvert}

\newcommand{\I}{\mathds{1}}

\newcommand{\setft}[1]{\mathrm{#1}}

\newcommand{\complex}{\mathbb{C}}

\renewcommand{\natural}{\mathbb{N}}
\newcommand{\integer}{\mathbb{Z}}

\newcommand\V{\mathcal{V}}
\newcommand\U{\mathcal{U}}

\renewcommand\H{\mathcal{H}}
\newcommand\J{\mathcal{J}}

\newcommand{\End}{\setft{End}}

\newcommand{\lip}{\langle}
\newcommand{\rip}{\rangle}
\newcommand{\braket}[2]{\lip #1\vert #2 \rip}

\newcommand{\ootimes}{ \otimes \cdots \otimes }

\newcommand{\bfd}{\mathbf{d}}

\newcommand{\frakS}{\mathfrak{S}}

\newcommand{\bfalpha}{\boldsymbol{\alpha}}
\newcommand{\bfbeta}{\boldsymbol{\beta}}

\newcommand{\eps}{\varepsilon}

\newcommand{\injects}{\hookrightarrow}

\def\ba#1\ea{\begin{align}#1\end{align}}

\begin{document}

\emergencystretch 3em
\title{\bf Constructive counterexamples to the additivity of minimum output Rényi entropy of quantum channels for all $p>1$
}

\author[$*\dagger$]{Harm Derksen}
 \author[$\ddagger$]{Benjamin Lovitz\thanks{emails: ha.derksen@northeastern.edu, benjamin.lovitz@concordia.ca}}
  \affil[$\dagger$]{Northeastern University, Boston, USA}
  \affil[$\ddagger$]{Concordia University, Montreal, Canada}

\date{\today}

\maketitle
\begin{abstract}
We present explicit quantum channels with strictly sub-additive minimum output Rényi entropy for all $p>1$, improving upon prior constructions which handled $p>2$. Our example is provided by explicit constructions of linear subspaces with high geometric measure of entanglement. This construction applies in both the bipartite and multipartite settings. As further applications, we use our construction to find entanglement witnesses with many highly negative eigenvalues, and to construct entangled mixed states that remain entangled after perturbation.
\end{abstract}

\section{Introduction}

Often, it is easier to prove that a randomly chosen object satisfies a given property than to exhibit a single, explicit object that satisfies that property. This is often referred to as the \textit{hay-in-a-haystack problem}, and has appeared in fields as diverse as number theory~\cite{khinchin1997continued}, algebraic geometry~\cite{landsberg2019towards}, circuit complexity~\cite{find2016better}, and coding theory~\cite{ta2017explicit}. In this paper, we address a hay-in-a-haystack problem for an additivity problem in quantum information theory.


As observed by Hayden and Winter for $p>1$~\cite{hayden2008counterexamples}, and by Hastings for $p=1$~\cite{hastings2009superadditivity} there exist randomized constructions of quantum channels $\Phi$ and $\Psi$ for which the minimum output Rényi $p$-entropy is strictly sub-additive, i.e.
\ba
H_{\min,p}(\Phi \otimes \Psi) < H_{\min,p}(\Phi)+H_{\min,p}(\Psi).
\ea
Derandomizing Hastings' construction is a major open problem, in part due to its equivalence to several other channel capacity problems such as superadditivity of the Holevo capacity~\cite{shor2004equivalence}. Despite considerable efforts across more than 15 years since these randomized constructions came to light, all of the known explicit constructions only handle $p>2$~\cite{grudka2010constructive,brannan2018highly,szczygielski2024new} or $p$ close to (or equal to) zero~\cite{cubitt2008counterexamples}. In this work, we give explicit constructions of quantum channels with strictly subadditive minimum output Rényi entropy for all $p>1$.

By associating a quantum channel $\Phi: \End (\H_{\text{in}}) \rightarrow \End (\H_A)$ with the image $\U$ of the isometry $U: \H_{\text{in}} \rightarrow \H_A \otimes \H_B$ appearing in its Stinespring representation, it is equivalent to exhibit linear subspaces $\U \subseteq \H_A \otimes \H_B$ and $\V \subseteq \H_{A'} \otimes \H_{B'}$ for which
\ba
H_{\min,p}(\U \otimes \V) < H_{\min,p}(\U)+H_{\min,p}(\V).
\ea
See e.g.~\cite{hayden2007maximal,grudka2010constructive}. Here, $H_{\min,p}(\U)$ denotes the minimum Rényi $p$-entropy of the squared Schmidt coefficients of any unit vector in $\U$, and we view $\U \otimes \V$ as a subspace of the bipartite space $(\H_A \otimes \H_{A'})\otimes (\H_B \otimes \H_{B'})$. We let $H_{\min}(\U):=H_{\min,1}(\U)$. Our main result is the following:

\begin{theorem}\label{thm:subadditivity}
For any $p>1$, there is an explicit subspace $\U \subseteq \complex^n \otimes \complex^n$ for which $H_{\min,p} (\U\otimes \U) < 2 H_{\min,p} (\U)$, where $n$ can be taken as $n=2^{\tilde{{O}}((p-1)^{-1})}$ as $p \rightarrow 1^+$.
\end{theorem}
Here, $\tilde{O}$ suppresses a factor of $-\log(p-1)$. For example, when $p=2$ we show that $n=28$ suffices. In this case, $\dim(\U)=574$. Translating to quantum channels, this construction yields an explicit quantum channel $\Phi :\End(\complex^{574})\rightarrow \End(\complex^{28})$ such that $H_{\min,2}(\Phi^{\otimes 2}) < 2 H_{\min,2}(\Phi)$. See Remark~\ref{rmk:loosely_optimized} for a loosely optimized table for particular values of $p$.


To prove this result, we employ a common reduction to the problem of exhibiting subspaces with high geometric measure of entanglement~\cite{hayden2007maximal,grudka2010constructive,brannan2018highly}. For a subspace $\U\subseteq \complex^{n_1}\ootimes \complex^{n_m}$, we let
\ba
E(\U):=1-\max_{\ket{\phi_i}\in \complex^{n_i}} \bra{\phi_1\ootimes \phi_m}\Pi_\U \ket{\phi_1\ootimes \phi_m}
\ea
be the \textit{geometric measure of entanglement} of $\U$. Here and throughout, $\Pi_{\U}$ denotes the (orthogonal) projection to $\U$, $\ket{\psi}$ denotes a unit vector, and $\psi$ denotes a (not necessarily normalized) vector. We construct subspaces of high geometric measure of entanglement in both the bipartite and multipartite settings (although the bipartite setting is all that is needed for the subadditivity problem). We say a subspace $\U$ is \textit{entangled} if it contains no product states, i.e. $E(\U) >0$.

Several \textit{randomized} constructions of subspaces with high min-entropy~\cite{hayden2006aspects,hastings2009superadditivity,brandao2010hastings,fukuda2010comments,fukuda2010entanglement,aubrun2011hastings,belinschi2012eigenvectors,aubrun2014entanglement} or high geometric measure of entanglement~\cite{harrow2013church,bhaskara2024new} have appeared in the literature. However, these random techniques provide little information on how to find explicit constructions. To quote~\cite{brannan2018highly}: ``...there is a need for a systematic development of non-random examples of highly entangled subspaces."

Focusing for simplicity on the case $n:=n_1=\dots=n_m$, we obtain the following. It is well-known that the maximum dimension of an entangled subspace is $n^m-m(n-1)-1$~\cite{Par04,harris2013algebraic}.

\begin{theorem}\label{thm:main} Let $\H=(\complex^n)^{\otimes m}$.
\begin{enumerate}
\item There is an explicit entangled subspace $\U \subseteq \H$ of maximal dimension $n^m-m(n-1)-1$ for which $E(\U) \geq m^{-(n-1)m}$.
\item For any $0< \eps < 1$, there is an explicit subspace $\U\subseteq \H$ for which $\dim(\U) = (1-o(1))(1-\eps) n^m$ and $E(\U)\geq {\varepsilon m^{-m}}$. In the special case when $\eps=\binom{md}{d,\dots,d}^{-1}$ for some positive integer $d$, this bound can be strengthened to $E(\U) \geq \eps$.
\end{enumerate}
\end{theorem}

Here and throughout, $o(1)$ denotes a function that approaches zero in the limit $n \rightarrow \infty$ when all other parameters are fixed. Part 1 of this theorem addresses Question 1 in~\cite{hayden2006aspects}. Part 2 of this theorem implies Theorem~\ref{thm:subadditivity}, using the well-known bounds $H_{\min,p}(\U) \geq \frac{1}{1-p} \log ( E(\U)^p + (1-E(\U))^p)$ and $H_{\min,p}(\U \otimes \overline{\U}) \leq \frac{p}{1-p} \log \left[\frac{\dim (\U)}{n^2}\right]$, along with a careful analysis of the $o(1)$ term appearing in the dimension of $\U$ (for our construction, $\U=\overline{\U}$). See Section~\ref{sec:construction}.

Entangled subspaces are also useful for certifying entanglement of mixed quantum states via the range criterion~\cite{Hor97,BDMSST99} and constructing entanglement witnesses~\cite{ATL11,CS14}. In Section~\ref{sec:applications} we illustrate these applications of our construction. In particular, we construct entangled mixed states that remain entangled after perturbation, and we construct entanglement witnesses with many, highly negative eigenvalues.

\subsection{Construction overview}

We now give a high-level overview of our construction. For simplicity, we focus on the bipartite setting $m=2$ with $n_1=n_2=n$. For a collection of non-zero scalars $C=(C_{i,j})_{i,j=1}^n$, let

\ba\label{eq:Uc}
\U_{C}:=\{\psi \in \complex^n \otimes \complex^n : \sum_{i+j=k} C_{i,j} \psi_{i,j}=0 \quad \text{for all} \quad k=2,3,\dots,2n\}.
\ea

First observe that $\U_{C}$ is an entangled subspace: Suppose toward contradiction that $\ket{\phi}:=\ket{\phi_1}\otimes \ket{\phi_2}$ is a product unit vector in $\U_C$. Let $\alpha_1, \alpha_2 \in [n]$ be the smallest indices for which $\phi_{1,\alpha_1}\neq 0$ and $\phi_{2,\alpha_2}\neq 0$, respectively, and let $k=\alpha_1+\alpha_2$. Then $\sum_{i+j=k} C_{i,j} \phi_{i,j}=C_{\alpha_1, \alpha_2} \phi_{1,\alpha_1} \phi_{2,\alpha_2} \neq 0$, a contradiction.

Note also that $\dim(\U_C)=(n-1)^2$, which is the maximum dimension of an entangled subspace~\cite{Par04} (see also~\cite[Definition 11.2]{harris2013algebraic}). One can verify that the antisymmetric subspace used in~\cite{grudka2010constructive} as well as the constructions of entangled subspaces used in~\cite{CMW08} are examples of subspaces of the form $\U_C$.

In this work, we exhibit a particular choice of $C$ for which $E(\U_C)$ is bounded away from zero. For example, we prove that the subspace

\ba\label{eq:U_intro}
\U:=\bigg\{\psi  : \sum_{\substack{i,j \in [n]\\i+j=k}} \binom{n-1}{i-1}^{1/2}\binom{n-1}{j-1}^{1/2} \psi_{i,j}=0 \quad \text{for all} \quad k=2,\dots,2n\bigg\} \subseteq (\complex^n)^{\otimes 2}
\ea
satisfies $E(\U) \geq 2^{-2n+2}$.

Our choice of coefficients is tailored so that we can invoke a result of~\cite{beauzamy1990products} on symmetric projections. The \textit{symmetric subspace} $S^d(\complex^a) \subseteq (\complex^a)^{\otimes d}$ is the subspace of tensors invariant under all permutations of the $d$ subsystems. It is a standard fact that the projection onto $S^d(\complex^a)$ is given by $\Pi_{a,d}:=\frac{1}{d!} \sum_{\sigma \in \frakS_d} U_{\sigma}$, where $U_{\sigma}$ is the unitary that permutes subsystems according to the permutation $\sigma$, i.e. $U_{\sigma}\ket{i_1\cdots i_d}=\ket{i_{\sigma^{-1}(1)}\cdots i_{\sigma^{-1}(d)}}$.

\begin{theorem}[\cite{beauzamy1990products}]\label{thm:beauzamy_intro}
For any unit vectors $\ket{\psi} \in S^{d}(\complex^a)$, $\ket{\phi} \in S^{d}(\complex^a)$, it holds that
\ba\label{eq:product_bound_intro}
\|\Pi_{a,2d}(\ket{\psi} \otimes \ket{\phi})\|\geq {2d\choose d}^{-1/2}.
\ea
\end{theorem}
We let $a=2$ and $d=n-1$ so that $\H_n:=S^{n-1}(\complex^2)$ is $n$-dimensional. As an immediate consequence of this theorem, $\U:=\ker(\Pi_{a,2d})$ is an entangled subspace of the bipartite space $\H_n \otimes \H_n$, and $E(\U) \geq \binom{2n-2}{n-1}^{-1}\geq 2^{-2n+2}$. It turns out that $\U$ is precisely the subspace defined in~\eqref{eq:U_intro} when the standard basis of $\complex^n$ is identified with a standard ({monomial}) orthonormal basis of $\H_n$.

We generalize this construction in several directions. First, we generalize to the case of potentially unequal subsystem dimensions as well as to the multipartite setting. We also generalize this construction to produce entangled subspaces for which $E(\U)$ is higher, at the expense of $\dim(\U)$ being smaller. To do this, we choose $d$ smaller (and $a$ larger) so that $S^d(\complex^a)$ is still (at least) $n$-dimensional. This makes the lower bound in~\eqref{eq:product_bound_intro} higher, at the expense of $\U=\ker(\Pi_{a,2d})$ having lower dimension. In coordinates, the resulting subspaces are no longer of the form $\U_C$, but they can still be described explicitly (see Section~\ref{sec:robust}).

\subsection{Related work}\label{sec:related_work}

In this section we compare our work to previous constructions of highly entangled subspaces. We consider explicit and randomized constructions separately.

\subsubsection{Explicit constructions}

Many constructions of highly entangled subspaces have been presented in the literature. An entangled subspace of maximum dimension was presented by Bhat~\cite{Bha06}. Later, maximum-dimensional subspaces avoiding the set of states of low Schmidt rank were presented in~\cite{CMW08}. The antisymmetric subspace is an entangled subspace, which was used in~\cite{grudka2010constructive} to construct quantum channels exhibiting subadditivity for $p>2$. We note that our subspace $\U$ is precisely the antisymmetric subspace of $\complex^a \otimes \complex^a$ when $d=1$. Further examples were recently presented in~\cite{szczygielski2024new}.

Explicit constructions of highly entangled subspaces were also presented in~\cite{brannan2018highly} in the bipartite setting using the representation theory of free orthogonal quantum groups. Various properties of an associated class of quantum channels are analyzed in~\cite{brannan2020temperley}, along with a more general class of channels they call \textit{Temperley-Lieb} channels. In more details, the subspace they construct is an irreducible subrepresentation of the tensor product of two representations of free orthogonal quantum groups. A similar idea is also used to construct entangled subspaces in~\cite{nuwairan2013potential,al2014extreme} using representations of $\setft{SU}(2)$, although they only prove bounds on the entropy in very special cases, such as when one of the spaces has dimension $2$~(see~\cite[Section 2.4]{nuwairan2013potential} and~\cite[Remark 6]{brannan2018highly}). 

We now compare in detail our construction to the one presented in~\cite{brannan2018highly}. For simplicity, we focus on the case $n_1=n_2=n$. In effect, the authors of~\cite{brannan2018highly} endeavour to construct a subspace $\U$ of dimension $\ell$ for which the quantity
\ba
\mu(\U):=\frac{H_{\min}(\U)}{\log (n^2/\ell)}
\ea
is as large as possible (see~\cite[Equation 1]{brannan2018highly}).
The authors construct a subspace for which $E(\U) \geq 1-(\frac{\ell}{n^2})^{1/2}-o(1)$. From here, it follows that
\ba
H_{\min}(\U) &\geq -\log (1-E(\U))\\
		& \geq \frac{1}{2}\log(\frac{n^2}{\ell})-o(1),
\ea
which implies the bound $\mu(\U)\geq \frac{1}{2}-o(1)$.
By comparison, our construction applies to both the bipartite and multipartite settings, and in the bipartite setting we produce subspaces with $\mu(\U)$ arbitrarily close to 1. Indeed, when $\eps=\binom{2d}{d}^{-1}$ our construction achieves $H_{\min}(\U) \geq - \log (1-\eps)$ with $\ell/n^2=(1-o(1))(1-\eps)$, so by increasing $d$, $\mu(\U)$ can be brought arbitrarily close to $1$.

\subsubsection{Randomized constructions}

Randomized constructions of subspaces of high min-$p$-entropy were given for all $p>1$ in~\cite{hayden2008counterexamples}. In particular, they gave a randomized construction of a subspace $\V \subseteq \complex^{n} \otimes \complex^{n}$ for which
\ba
2 H_{\min,p}(\V) - H_{\min,p}(\V \otimes \overline{\V}) = \log(n)-\mathcal{O}(1).
\ea
By contrast, our subspace $\U$ can achieve at most a constant gap because it contains elements of Schmidt rank 2 (see Remark~\ref{rmk:rank2}), hence $H_{\min,p}(\U)\leq  1$.

Relatedly, there have been several randomized constructions of subspaces with high min-entropy $H_{\min}(\U)$~\cite{hayden2006aspects,hastings2009superadditivity,brandao2010hastings,fukuda2010comments,fukuda2010entanglement,aubrun2011hastings}. Hastings~\cite{hastings2009superadditivity} used such a randomized construction to exhibit an explicit subspace $\V \subseteq \complex^{n_1}\otimes \complex^{n_2}$ for which
\ba
H_{\min}(\V \otimes \overline{\V}) < 2 H_{\min}(\V).
\ea
The works~\cite{brandao2010hastings,fukuda2010comments,fukuda2010entanglement,aubrun2011hastings} give alternate proofs of Hastings' result. A common approach is to construct a random subspace $\V \subseteq \complex^{n_1} \otimes \complex^{n_2}$ with $n_2= \alpha n_1^2$ and $\dim(\U) = \beta n_2$ (for some constants $\alpha,\beta$) such that $H_{\min}(\V) \geq \log(n_1)- O(1/{n_1})$. The desired inequality then holds for $n_1 \gg 0$ by the bound
\ba
H_{\min}(\V \otimes \overline{\V}) &\leq (1-\frac{\dim(\V)}{n_1n_2}) \log(n_1^2-1)+h(\frac{\dim(\V)}{n_1n_2})\label{eq:VVlb}\\
 &\leq 2(1-\frac{\beta}{\alpha n_1}) \log(n_1)+h(\frac{\beta }{ \alpha n_1})\\
&=2 \log(n_1) - \Omega(\frac{\log n_1}{n_1}),
\ea
where $h(x):=-x \log x - (1-x) \log (1-x)$ is the binary entropy. The first inequality can be found in \cite[Theorem 5]{fukuda2010entanglement}, and holds whenever $\dim(\U) n_1 \geq n_2$ (see also~\cite[Equation 10]{hastings2009superadditivity} and~\cite[Lemma B.1]{brandao2010hastings}). While it is still possible that our subspace $\U$ exhibits strict subadditivity of MOE, it would require a different proof technique since our $\U$ contains elements of Schmidt rank 2 (see Remark~\ref{rmk:rank2}), and hence $H_{\min}(\U)\leq  1$.

See also~\cite{nema2022approximate} for partial progress on derandomizing Hastings' example using unitary designs.

Randomized constructions of subspaces of high geometric measure of entanglement are given in~\cite{harrow2013church}. For simplicity we focus on the case $m=2$, $n_1=n_2=n$. The more general setting can be analyzed similarly. In particular, the following randomized construction is given in~\cite{harrow2013church}:
\begin{prop}[Proposition 11 in~\cite{harrow2013church}]
Let $\U \subseteq \complex^n \otimes \complex^n$ be a Haar-random subspace of dimension $\ell=n^2-2(n-1)-x$ for some positive integer $x$. Then
\ba
\Pr(E(\U) \leq n^{-2n/x-2}) \leq n^{-n}.
\ea
\end{prop}
We now compare the bound on $E(\U)$ obtained by this randomized construction to the one obtained by our explicit construction. We consider two examples $\ell=(n-1)^2$ (the largest possible) and $\ell=(1-o(1))(1-\eps)n^2$. We find that our explicit construction outperforms the randomized one in both cases:
\begin{itemize}
\item When $\ell=(n-1)^2$, the randomized construction obtains $E(\U) > n^{-2n-2}$. By comparison, our explicit construction obtains the better bound $E(\U)\geq 2^{-2n+2}$.
\item When $\ell=(1-o(1))(1-\eps)n^2$, the randomized construction obtains
\ba
E(\U) > n^{{-2n}/{(\eps n^2 -2n+2+o(1)(1-\eps)n^2)}-2}.
\ea
Note that this bound is less than $n^{-2}$ for $n \gg0$. By comparison, our explicit construction obtains a constant bound $E(\U) \geq \eps/4$ (or $E(\U) \geq \eps$ when $\eps=\binom{2d}{d}^{-1}$ for some $d$). We note that the construction of~\cite{brannan2018highly} also achieves a constant bound of $E(\U) \geq 1- \sqrt{1-\eps}$.
\end{itemize}

Randomized constructions of highly entangled subspaces are also presented in~\cite{bhaskara2024new} over the real numbers. In particular, they construct a subspace of dimension $\dim(\U)=c n^2$ for a fixed constant $c$, with $E(\U)$ depending inverse-polynomially in $n$. See~\cite[Theorem 7.1]{bhaskara2024new} for a more general statement, which holds for other types of entanglement as well as in the smoothed analysis setting.


\section{The construction}\label{sec:construction}

In this section we prove Theorem~\ref{thm:subadditivity}. Our starting point is the following, which is essentially just a rephrasing of Theorem~\ref{thm:beauzamy_intro}.

\begin{theorem}\label{thm:sub_for_subadditivity}
Let $\U= S^{2d}(\complex^a)^{\perp} \subseteq S^d(\complex^a) \otimes S^d(\complex^a)$, where $S^d(\complex^a) \otimes S^d(\complex^a)$ is regarded as a bipartite Hilbert space with local Hilbert spaces $\H_A=\H_B=S^d(\complex^a)$. Then $\U=\overline{\U}$ and $E(\U) \geq \binom{2d}{d}^{-1}$.
\end{theorem}

\begin{proof}
It is clear that $\U=\overline{\U}$ because $S^{2d}(\complex^a)$ has a real basis. Furthermore,
\ba
E(\U)&= \min_{\substack{\ket{\psi}\in \U\\ \ket{\phi_1},\ket{\phi_2}\in S^d(\complex^a) }} {(1-|\braket{\psi}{\phi_1\otimes \phi_2}|^2)}\\
&= {\min_{\ket{\phi_1},\ket{\phi_2}\in S^d(\complex^a)}{\norm{\Pi_{a, 2d} \ket{\phi_1 \otimes \phi_2}}}}^2\\
&\geq {2d \choose d}^{-1},
\ea
where the last line follows from Theorem~\ref{thm:beauzamy_intro}. This completes the proof.
\end{proof}

In fact, by~\cite[Theorem 3]{beauzamy1996massively}, the inequality~\eqref{eq:product_bound_intro} appearing in Theorem~\ref{thm:beauzamy_intro} is sharp, and hence $E(\U)=\binom{2d}{d}^{-1}$.

\subsection{Rényi entropy bounds}
For a unit vector $\ket{\psi} \in \H_A \otimes \H_B$, and a constant $p>1$, let
\ba
H_{p}(\psi)=\frac{1}{1-p} \log(\sum_{i} \lambda_i^p)
\ea
be the \textit{Rényi $p$-entropy} of $\psi$, where the logarithm is taken base-2, and $\lambda_i$ are the squared Schmidt coefficients of $\ket{\psi}$. For a subspace $\U \subseteq \H_A \otimes \H_B$, define the \textit{min-$p$-entropy} $H_{\min,p}(\U)$ to be the minimum Rényi $p$-entropy of any unit vector in $\U$. The following lemma is well-known, see e.g.~\cite[Equation 26]{hayden2007maximal},~\cite[Lemma 1]{grudka2010constructive}, or~\cite[Lemmas 2 and 3]{szczygielski2024new}. We prove it here for completeness.

\begin{lemma}
If $\U \subseteq \H_A \otimes \H_B$ is a subspace, then $\U \otimes \overline{\U} \subseteq (\H_A \otimes \H_{A'}) \otimes (\H_B \otimes \H_{B'})$ satisfies
\ba
H_{\min,p}(\U \otimes \overline{\U}) \leq \frac{p}{1-p} \log \left[\frac{\dim (\U)}{\dim(\H_A) \dim(\H_B)}\right].
\ea
\end{lemma}
In the statement of the lemma, we view $\U \otimes \overline{\U}$ as a subspace of bipartite space according to the bipartition $(\H_A \otimes \H_{A'}) \otimes (\H_B \otimes \H_{B'})$.

We also require the following lower bound on $H_{\min,p}(\U)$. This bound can be proven easily using Schur-concavity of the Rényi $p$-entropy. See also~\cite[Lemma 4]{szczygielski2024new}. We prove it here because it is slightly less standard than the above upper bound.

\begin{lemma}
If $\dim(\H_A), \dim(\H_B) \geq 2$, $\U \subseteq \H_A \otimes \H_B$ is a subspace, and $\varepsilon$ satisfies $E(\U) \geq \varepsilon \geq 0$, then
\ba
H_{\min,p}(\U) \geq \frac{1}{1-p} \log ( \varepsilon^p + (1-\varepsilon)^p).
\ea
\end{lemma}
\begin{proof}
Let $n=\min\{\dim(\H_A),\dim(\H_B)\}$, let $\psi \in \U$ be any unit vector, and let $\lambda_1 \geq \dots \geq \lambda_{n}$ be the squared Schmidt coefficients of $\psi$. Then $\lambda_1 \leq 1-\varepsilon$, so $(1-\varepsilon, \varepsilon,0,\dots, 0)$ majorizes $(\lambda_1,\dots, \lambda_n)$. By Schur-concavity of the Renyi entropy,
\ba
H_{p}(\psi) \geq H_p(1-\varepsilon, \varepsilon,0,\dots, 0)=\frac{1}{1-p} \log ( \varepsilon^p + (1-\varepsilon)^p).
\ea
This completes the proof.
\end{proof}

\subsection{Proof of Theorem~\ref{thm:subadditivity}}


\begin{theorem}[Theorem~\ref{thm:subadditivity} in more details]\label{thm:subadditivity_details}
For any $p >1$, there exists a choice of $a,d$ for which the subspace $\U=\overline{\U}=S^{2d}(\complex^a)^{\perp} \subseteq S^d(\complex^a)\otimes S^d(\complex^a)$ satisfies $H_{\min,p}(\U\otimes \U) < 2H_{\min,p}(\U)$. Furthermore, under this choice it holds that  $n:=\dim(S^d(\complex^a))=2^{\mathcal{O}(-(p-1)^{-1}\log(p-1))}$.
\end{theorem}
\begin{proof}
By the above bounds, it suffices to verify that
\ba
\frac{p}{1-p} \log \left[\frac{\dim (\U)}{n^2}\right] < 2\frac{1}{1-p} \log ( \varepsilon^p + (1-\varepsilon)^p),
\ea
for some $\varepsilon$ satisfying $E(\U) \geq \varepsilon \geq 0$. Equivalently, we need to show that
\ba\label{eq:firstgoal}
\frac{\dim (\U)}{n^2} >(\varepsilon^p + (1-\varepsilon)^p)^{2/p}.
\ea
By Theorem~\ref{thm:sub_for_subadditivity}, $E(\U) \geq \varepsilon (d):= \binom{2d}{d}^{-1}$. Note that $\dim(\U)=n^2-\binom{a+2d-1}{2d}$. Hence
\ba
\frac{\dim(\U)}{n^2}=1-\frac{\binom{a+2d-1}{2d}}{\binom{a+d-1}{d}^2}=1-\varepsilon(d)\prod_{i=0}^{d-1}\left(1+\frac{d}{a+i}\right),
\ea
which approaches $1-\varepsilon(d)$ as $a \rightarrow \infty$. It is clear that for $d$ large enough, it holds that
\ba
1-\varepsilon(d) > (\varepsilon(d)^p+(1-\varepsilon(d))^p)^{2/p}.
\ea
Hence, for any fixed $p>1$ there exists a choice of $a$ and $d$ for which strict subadditivity holds.

We now find a particular choice of $a=a(p)$ and $d=d(p)$ for which strict subadditivity holds. Let $\varepsilon=\varepsilon(d)$, and let
\ba
d=\biggceil{\frac{10}{p-1}},\quad a=5 d^2.
\ea
Note that
\ba
\prod_{i=0}^{d-1}\left(1+\frac{d}{a+i}\right)\leq\left(1+\frac{d}{a}\right)^d \leq  e^{d^2/a}=e^{\frac{1}{5}},
\ea
where the inequality follows from $(1+x) \leq e^x$ for any real number $x$.
So it suffices to show that
\ba\label{eq:goal}
1-\varepsilon \; e^{1/5} > (\epsilon^p+(1-\epsilon)^p)^{2/p}.
\ea
When $p \geq 11$ we have $d=1$ and this inequality is easily verified. Assume from now on that $p < 11$, so $d \geq 2$, and hence $0<\varepsilon\le \frac16$.

\medskip\noindent
First we factor $(1-\varepsilon)^p$:
\[
\varepsilon^p+(1-\varepsilon)^p
=(1-\varepsilon)^p\Bigl(1+\Bigl(\frac{\varepsilon}{1-\varepsilon}\Bigr)^p\Bigr).
\]
Let $u=\bigl(\frac{\varepsilon}{1-\varepsilon}\bigr)^p$. Since $\varepsilon\le \frac16$ we have
$\frac{\varepsilon}{1-\varepsilon}\le \frac15$, hence $0<u<1$. Then
\[
\bigl(\varepsilon^p+(1-\varepsilon)^p\bigr)^{2/p}
=(1-\varepsilon)^2(1+u)^{2/p}.
\]
Because $p>1$ we have $2/p<2$, and since $1+u>1$,
\[
(1+u)^{2/p}<(1+u)^2=1+2u+u^2<1+3u
\qquad(0<u<1).
\]
Thus
\[
\bigl(\varepsilon^p+(1-\varepsilon)^p\bigr)^{2/p}
< (1-\varepsilon)^2(1+3u)
=(1-\varepsilon)^2+3(1-\varepsilon)^2u.
\]
Now
\[
(1-\varepsilon)^2u=(1-\varepsilon)^2\Bigl(\frac{\varepsilon}{1-\varepsilon}\Bigr)^p
=\varepsilon^p(1-\varepsilon)^{2-p}.
\]
Since $p<11$, we have $2-p>-9$, hence $(1-\varepsilon)^{2-p}\le (1-\varepsilon)^{-9}\le (6/5)^9<6$. Therefore
\[
3\,\varepsilon^p(1-\varepsilon)^{2-p}<18\,\varepsilon^p,
\]
and hence
\begin{equation}\label{eq:rhs-bound}
\bigl(\varepsilon^p+(1-\varepsilon)^p\bigr)^{2/p}
< (1-\varepsilon)^2+18\,\varepsilon^p
=1-2\varepsilon+\varepsilon^2+18\varepsilon^p.
\end{equation}
So \eqref{eq:goal} follows if we show
\begin{equation}\label{eq:key}
(2-e^{1/5})\varepsilon>\varepsilon^2+18\varepsilon^p.
\end{equation}
Dividing by $\varepsilon$, it suffices to prove
\ba\label{eq:key}
2-e^{1/5}>\varepsilon+18\varepsilon^{p-1}.
\ea
Now,
\ba
\varepsilon^{p-1} \leq \varepsilon^{\frac{10}{d}}\leq \frac{(2d+1)^{\frac{10}{d}}}{4^{10}} \leq \frac{5^5}{4^{10}}< \frac{1}{300}.
\ea
In the second inequality we used $\binom{2d}{d} \geq \frac{4^d}{2d+1}$, and in the second inequality we used the fact that the function $x \mapsto (2x+1)^{1/x}$ is decreasing for $x \geq 2$ (which in turn can be verified by showing that the logarithm $x\mapsto \frac{\ln(2x+1)}{x}$ is decreasing for $x \geq 2$). Hence,
\ba
\varepsilon+18\varepsilon^{p-1} < \varepsilon + \frac{3}{50}.
\ea
From here, the strict inequality~\eqref{eq:key} follows easily since $\varepsilon \leq 1/2$. Note that for this construction we have $n=\binom{a+d-1}{d} \leq (a+d-1)^d \leq 2^{\mathcal{O}(-(p-1)^{-1} \log(p-1))}$. This completes the proof.
\end{proof}

\begin{remark}\label{rmk:loosely_optimized}
For particular values of $p$ we can optimize $d$ and $a$ so that $n=\binom{a+d-1}{d}$ is as small as possible subject to the inequality~\eqref{eq:firstgoal} holding. We record a loosely optimized table of sufficient values for $d$ and $a$, along with $n=\binom{a+d-1}{d}$ and $\dim(\U)=n^2-\binom{a+2d-1}{2d}$, for a few values of $p$:

\begin{center}
\begin{tabular}{||c c c c c ||}
 \hline
 $p$ & $d$ & $a$ & $n$ & $\dim(\U)$ \\ [0.5ex]
 \hline\hline
 2  & 2 & 7 & 28 & 574 \\
 \hline
 $1.5$  & 2 &13&91 & 6461 \\
 \hline
 $1.25$   & 3 &44&15,180& $\approx 2.16 \cdot 10^8$ \\
 \hline
 $1.125$  & 5 &240 & $\approx 6.92 \cdot 10^9$ & $\approx 4.76 \cdot 10^{19}$ \\
 \hline
  $1.0625$  & 10 &889&$\approx 8.94 \cdot 10^{22}$ & $\approx 7.99 \cdot 10^{45}$  \\
  \hline
\end{tabular}
\end{center}
\end{remark}

\section{Constructing entangled subspaces in other parameter regimes}\label{sec:robust}

In this section we explicitly construct subspaces of $\H=\complex^{n_1}\otimes \dots \otimes \complex^{n_m}$ with high geometric measure of entanglement. Let $S= [n_1] \times \dots \times [n_m]$. For a tuple of coefficients $C=(C_{\bfalpha})_{\bfalpha \in S}$, let
\ba\label{eq:uc}
\U_{C}:=\bigg\{\psi \in \H : \sum_{\substack{\bfalpha \in S\\ |\bfalpha|=k}}C_{\bfalpha} \; \psi_{\bfalpha}=0\quad\text{for all}\quad k \bigg\},
\ea
where $|\bfalpha|:=\alpha_1+\dots+\alpha_m$, and $k$ ranges over $\{m, m+1, \dots, n_1+\dots + n_m\}$.
\begin{prop}
If $C_{\bfalpha} \neq 0$ for all $\bfalpha\in S$, then $\U_C$ is a completely entangled subspace (i.e. it contains no product states) of maximum dimension.
\end{prop}
\begin{proof}
Let $\ket{\phi}=\ket{\phi_1 \otimes \dots \otimes \phi_m}$ be a product state. For each $i\in [m]$ let $\alpha_i$ be the smallest index for which $\phi_{i, \alpha_i} \neq 0$, and let $k=\sum_i \alpha_i$. Then
\ba
\sum_{\substack{\bfalpha \in S\\ |\bfalpha|=k}}C_{\bfalpha} \; \phi_{\bfalpha}=C_{\bfalpha} \phi_{1,\alpha_1} \cdots \phi_{m, \alpha_m}\neq 0,
\ea
so $\phi \notin \U_C$. This shows that $\U_C$ is completely entangled. Furthermore, $\dim(\U_C)=n_1\cdots n_m-\sum_{i=1}^m (n_i)+m-1$, which is the maximum dimension of a completely entangled subspace by~\cite{Par04} (see also~\cite[Definition 11.2]{harris2013algebraic}).
\end{proof}

More generally, let $P=(P_1,\dots, P_{\ell})$ be any ordered partition of $S$ that respects the ordering of each $[n_i]$ in the sense that if $\bfalpha \in P_{q}$ and $\gamma_i<\alpha_i$ then $(\alpha_1,\dots, \alpha_{i-1},\gamma_i,\alpha_{i+1},\dots, \alpha_m) \in P_r$ for some $r<q$. Let $\U_{C,P}$ be the set of $\psi$ for which $\sum_{\bfalpha \in P_i} C_{\bfalpha} \psi_{\bfalpha}=0$ for all $i$. Then $\U_{C,P}$ is completely entangled (but may not be of maximum dimension).

We exhibit choices of $C$ and $P$ for which $E(\U_{C,P})$ is high. For convenience, we summarize our main results in the case $n_1=\dots=n_m=n$:
\begin{theorem}\label{thm:nodetails} Let $\H=(\complex^n)^{\otimes m}$.
\begin{enumerate}
\item There is an explicit choice of $C\in \complex^S$ for which $E(\U_C) \geq m^{-(n-1)m}$ (and by construction, $\dim(\U_C)$ is maximal among all completely entangled subspaces).
\item For any $0< \eps < 1$, there is an explicit choice of $C$ and $P$ for which $\dim(\U_{C,P}) = (1-o(1))(1-\eps) n^m$ and $E(\U_{C,P})\geq {\varepsilon m^{-m}}$. In the special case when $\eps=\binom{md}{d,\dots,d}^{-1}$ for some positive integer $d$, this bound can be strengthened to $E(\U_{C,P}) \geq \eps$.
\end{enumerate}
\end{theorem}

\subsection{Choosing $C$ and $P$ for the construction}
We now explain how we choose $C$ and $P$ so that $U_{C,P}$ has high geometric measure of entanglement. We first choose positive integers $a, d_i$ with $n_i \leq \binom{a+d_i-1}{d_i}$, and view $\complex^{n_i}$ as a subspace of $S^{d_i}(\complex^{a})$ by sending the standard basis of $\complex^{n_i}$ to the the first $n_i$ elements of the orthonormal basis
\ba
\bigg\{\ket{\bfalpha_i}:=\binom{d_i}{\bfalpha_i}^{1/2} \Pi_{S^{d_i}(\complex^a)} (\ket{1}^{\otimes \alpha_{i,1}}\otimes \cdots \otimes \ket{a}^{\otimes \alpha_{i,a}}) : \bfalpha_i \in \{0,1,\dots, d_i\}^a,  |\bfalpha_i|=d_i \bigg\}
\ea
for $S^{d_i}(\complex^{a})$, where $\Pi_{S^{d_i}(\complex^a)}$ is the orthogonal projection onto  $S^{d_i}(\complex^{a})$. (Alternatively, any isometric embedding $\complex^{n_i} \injects S^{d_i}(\complex^a)$ will do.) Let $\bfd:=(d_1,\dots, d_m)$, and let $\J_{a,\bfd}:=S^{d_1}(\complex^a)\ootimes S^{d_m}(\complex^a)$.

After making the choice, we let $\Pi_{a,\bfd}$ be the orthogonal projection onto the symmetric subspace $S^{d_1+\dots+d_m}(\complex^a)\subseteq \J_{a,\bfd}$, and define
\ba
\U_{a,\bfd}:=\ker(\Pi_{a,\bfd})\subseteq \J_{a,\bfd}.
\ea
Then $\U_{a,\bfd}$ is a completely entangled subspace of $\H$ (after intersecting with $\H$ if the inequality $n_i \leq \binom{a+d_i-1}{d_i}$ is strict for some $i$). This can be verified from the following remark, or from Corollary~\ref{cor:entangled} below.

\begin{remark}[Writing $\U_{a,\bfd}$ in coordinates]
In coordinates, we can write $\U_{a,\bfd}$ as a subspace of the form $\U_{C,P}$ as follows. For each $\bfalpha = (\bfalpha_1,\dots, \bfalpha_m) \in \integer_{\geq 0}^{a \times m}$ with $(1,\dots, 1)\bfalpha = \bfd$
we let
\ba
C_{\bfalpha}:=\binom{d_1}{\bfalpha_1}^{1/2}\cdots \binom{d_m}{\bfalpha_m}^{1/2}.
\ea
Then
\ba\label{eq_bfbeta}
\U_{a,\bfd}= \bigg\{\psi : \sum_{\substack{\bfalpha \in \integer_{\geq 0}^{a \times m}\\ (1,\dots, 1)\bfalpha = \bfd\\ (1,\dots, 1) \bfalpha^\t= \bfbeta}} C_{\bfalpha} \psi_{\bfalpha} =0 \quad\text{for all}\quad \bfbeta   \bigg\}\subseteq \J_{a,\bfd},
\ea
where $\bfbeta=(\beta_1,\dots, \beta_a)$ ranges over $\{\bfbeta \in \integer_{\geq 0}^a : |\bfbeta|=|\bfd|\}$. One can verify that the expressions~\eqref{eq_bfbeta} and $\U_{a,\bfd}=\ker(\Pi_{a,\bfd})$ are equivalent by noting that $\ket{\psi}\in \ker(\Pi_{a,\bfd})$ if and only if
\ba
(\bra{1}^{\otimes \beta_1} \ootimes \bra{a}^{\otimes \beta_a})\Pi_{a,\bfd}\ket{\psi}=0
\ea
for all $\bfbeta$, and
\ba
(\bra{1}^{\otimes \beta_1} \ootimes \bra{a}^{\otimes \beta_a})\Pi_{a,\bfd}=\binom{|\bfd|}{\bfbeta}^{-1}\sum_{\substack{\bfalpha=(\bfalpha_1,\dots,\bfalpha_m) \in \integer_{\geq 0}^{a \times m}\\ (1,\dots, 1)\bfalpha = \bfd\\ (1,\dots, 1) \bfalpha^\t= \bfbeta}} C_{\bfalpha}  (\bra{\bfalpha_1}\ootimes \bra{\bfalpha_m}),
\ea
which is a tedious but elementary computation.
\end{remark}


 If $d_1=\dots=d_m=:d$ we let $\J_{a,d}=\J_{a,\bfd}$ and $\U_{a,d}=\U_{a,\bfd}$. Let us now restate Theorem~\ref{thm:nodetails} in more details.

\begin{theorem}\label{thm:details}
Let $\H=(\complex^n)^{\otimes m}$, and for $a,d$ with $n \leq \binom{a+d-1}{d}$ let $\U_{a,d} \subseteq \J_{a,d}$ be defined as above.
\begin{enumerate}
\item If $a=2, d=n-1$, then $\U_{a,d}\subseteq \H$ is completely entangled of maximum dimension, and $E(\U) \geq m^{-(n-1)m}$.
\item For any $0 < \eps < 1$, there is a choice of $d=d(\eps)$ (given by~\eqref{eq:kdepsilon}) for which letting $a=\lceil \sqrt[d]{d!\cdot n}\rceil$ gives $\dim(\U_{a,d}) = (1-o(1))(1-\eps) n^m$ and $E(\U_{a,d})\geq {\varepsilon m^{-m}}$. In the special case when $\eps=\binom{md'}{d',\dots,d'}^{-1}$ for some positive integer $d'$, one can choose $d=d'$ and this bound can be strengthened to $E(\U_{a,d}) \geq \eps$.
\end{enumerate}
\end{theorem}


\begin{example}[Bipartite setting]
In the bipartite setting,
\ba
\U_{2,n-1}=\bigg\{\psi  : \sum_{\substack{i,j \in [n]\\i+j=k}} \binom{n-1}{i-1}^{1/2}\binom{n-1}{j-1}^{1/2} \psi_{i,j}=0 \quad \text{for all} \quad k=2,\dots,2n\bigg\} \subseteq (\complex^n)^{\otimes 2}.
\ea
By Theorem~\ref{thm:details}, $\U_{2,n-1}$ is completely entangled of maximum dimension, and $E(\U) \geq 2^{-2n+2}$.
\end{example}

\subsection{Analyzing the geometric measure of entanglement for the construction}

In this section we prove Theorem~\ref{thm:details}, along with more general statements that handle the case of potentially unequal subsystem dimensions. We carry over the definitions from the previous section, including $\H=\complex^{n_1}\otimes \dots \otimes \complex^{n_m}$.

The starting point in our analysis is the following result of \cite{beauzamy1990products}.

\begin{theorem}\label{thm:beauzamy}
If $\psi \in S^{d_1}(\complex^a)$, $\phi \in S^{d_2}(\complex^a)$ and $\bfd=(d_1,d_2)$, then
$$
\|\Pi_{a,\bfd}(\psi \otimes \phi)\|\geq {d_1+d_2\choose d_1}^{-1/2}\|\psi\|\|\phi\|.
$$
\end{theorem}
A one-page proof of this result can be found in~\cite{zeilberger1994chu}. In these references, the result is stated as a bound on the Bombieri norm of the product of homogeneous polynomials. One can translate this to the above statement by identifying $\ket{\bfalpha_i}$ with the polynomial $\binom{d}{\bfalpha}^{1/2} x_1^{\alpha_1} \cdots x_a^{\alpha_a}$ and verifying that this map is a linear isometry for which $\Pi_{a,\bfd}(\psi \otimes \phi)$ corresponds to the product of $\psi$ and $\phi$ as polynomials.

\begin{cor}\label{cor:entangled}
If $\psi_i\in S^{d_i}(\complex^a)$ for $i=1,2,\dots,m$, and $\bfd=(d_1,\dots, d_m)$, then
$$
\|\Pi_{a, \bfd}(\psi_1\otimes \psi_2\otimes \cdots \otimes \psi_m)\|\geq {d_1+d_2+\cdots+d_m\choose d_1,d_2,\dots,d_m}^{-1/2}\|\psi_1\|\|\psi_2\|\cdots \|\psi_m\|.
$$
\end{cor}
\begin{proof}
We have
\ba
\|\Pi_{a, \bfd}(\psi_1\otimes \psi_2\otimes \cdots \otimes \psi_m)\|&=\|\Pi_{a,\bfd}(\psi_1\otimes \Pi_{a, (d_2,\dots,d_m)}(\psi_2\otimes \cdots \otimes \psi_m))\|\\
&\geq \binom{d_1+\dots+d_m}{d_1}^{-1/2} \|\psi_1\| \;\; \|\Pi_{a,(d_2,\dots,d_m)}(\psi_2\otimes \cdots \otimes \psi_m)\|\\
&\geq \dots \geq {d_1+d_2+\cdots+d_m\choose d_1,d_2,\dots,d_m}^{-1/2}\|\psi_1\|\;\;\|\psi_2\|\cdots \|\psi_m\|,
\ea
completing the proof.
\end{proof}

\begin{theorem}\label{theo:rank1angle}
Let $a,d_1,\dots, d_m, \ell$ be positive integers. If
 $n_i\leq {a+d_i-1\choose a-1}$ and
 \ba
 \ell \leq n_1 \cdots n_m - \dim(S^{d_1+\dots + d_m}(\complex^a))
 \ea
 then there exists a subspace $\U\subseteq \H$ of dimension $\ell$ with $E(\U)\geq {d_1+d_2+\cdots+d_m\choose d_1,d_2,\dots,d_m}^{-1}$.
\end{theorem}

\begin{proof}
Let $\U=\ker(\Pi_{a,\bfd}) \cap \H \subseteq \H$ (recall that we view $\complex^{n_i}$ as a subspace of $S^{d_i}(\complex^a)$), which has dimension at least $n_1 \cdots n_m - \dim(S^{d_1+\dots + d_m}(\complex^a))$. It remains to prove that $E(\U)\geq {d_1+d_2+\cdots+d_m\choose d_1,d_2,\dots,d_m}^{-1}$. Note that
\ba
E(\U)&= \min_{\substack{\ket{\psi}\in \U\\ \ket{\phi_i}\in \complex^{n_i}}} {(1-|\braket{\psi}{\phi_1\ootimes \phi_m}|^2)}\\
&\geq \min_{\substack{\ket{\psi}\in \ker(\Pi_{a,\bfd}) \\ \ket{\phi_i}\in S^{d_i}(\complex^{a_i})}} {(1-|\braket{\psi}{\phi_1\ootimes \phi_m}|^2)}\\
&= {\min_{\ket{\phi_i} \in S^{d_i}(\complex^{a_i})}{\norm{\Pi_{a, \bfd} \ket{\phi_1 \ootimes \phi_m}}}}^2\\
&\geq {d_1+d_2+\cdots+d_m\choose d_1,d_2,\dots,d_m}^{-1},
\ea
where the second line follows from $\U \subseteq \ker(\Pi_{a,\bfd})$ and $\complex^{n_i} \subseteq S^{d_i}(\complex^{a_i})$. This completes the proof.
\end{proof}

\begin{cor}\label{cor:maxnorank1}
There exists a subspace $\U\subseteq \H$ of dimension
\ba
\ell=n_1 \cdots n_m - \sum_{i=1}^m(n_i)+m-1
\ea
with $E(\U)\geq {n_1+n_2+\cdots+n_m-m\choose n_1-1,n_2-1,\dots,n_m-1}^{-1}$.
\end{cor}
Note that the subspace $\U$ constructed in this corollary is entangled of maximum dimension.
\begin{proof}
Take $a=2$, $d_i=n_i-1$ for $i=1,2,\dots,m$, and
\ba
\ell=n_1 \cdots n_m-{d_1+d_2+\cdots +d_m+a-1\choose a-1}=n_1 \cdots n_m - \sum_{i=1}^m(n_i)+m-1
\ea
in Theorem~\ref{theo:rank1angle}.
\end{proof}

Let us now look at the special case where $n_1=n_2=\cdots=n_m=n$.
\begin{cor}\label{cor:maxnorank1simple}
There exists a subspace $\U\subseteq (\complex^n)^{\otimes m}$ of dimension $n^m-mn+m-1$ with $E(\U)\geq m^{-(n-1)m}$.
\end{cor}
\begin{proof}
We take $n_1=n_2=\cdots=n_m=n$ in Corollary~\ref{cor:maxnorank1}.
Then $\U$ is a subspace of the specified dimension, and
$E(\U) \geq {m(n-1)\choose n-1,n-1,\dots,n-1}^{-1}\geq m^{-(n-1)m}$.
\end{proof}

We constructed entangled subspaces $\U$ of $(\complex^n)^{\otimes m}$ of maximal dimension
for which $E(\U)$ has a lower bound that is exponentially small in $n$. For subspaces of smaller dimension we can get better bounds. For example, we can construct $\U$ of dimension $(1-o(1))(1-\varepsilon) n^m$ for which $E(\U)$ is lower bounded by a constant independent of $n$.
\begin{prop}\label{prop:D1bound}
For any $d \in \natural$ there exists a subspace $\U\subseteq (\complex^n)^{\otimes m}$ of dimension
$$\ell \geq n^m- n^m{md\choose d,d,\dots,d}^{-1}\left(1+\frac{m e}{\sqrt[d]{n}}\right)^{m d}
$$
for which $E(\U) \geq {m d\choose d,d,\dots,d}^{-1}$.
\end{prop}
In the statement of the proposition, $e$ is Euler's constant.

\begin{proof}
In Theorem~\ref{theo:rank1angle} we will take $d_1=d_2=\cdots =d_m=d$ and $n_1=n_2=\cdots=n_m=n$.
We take $a=\lceil \sqrt[d]{d!\cdot n}\rceil\leq \sqrt[d]{d!\cdot n}+1$, so that
$${d+a-1\choose a-1}={d+a-1\choose d}\geq \frac{a^d}{d!}\geq n.
$$
Now we have
\begin{multline*}
{d_1+d_2+\cdots+d_m+a-1\choose a-1}={md+a-1\choose md}\leq \frac{(md+a-1)^{md}}{(md)!}\leq \frac{(md+\sqrt[d]{d!\cdot n})^{md}}{(md)!}=\\=n^m{md\choose d,d,\dots,d}^{-1}
\Big(1+\frac{md}{\sqrt[d]{d!\cdot n}} \Big)^{md}\leq n^m{md\choose d,d,\dots,d}^{-1}\Big(1+\frac{me}{\sqrt[d]{n}}\Big)^{md}.
\end{multline*}
This completes the proof.
\end{proof}

\begin{theorem}\label{maxnorank1smalldim}
    Let $0<\varepsilon<1$ be arbitrary but fixed. Then there exists a subspace $\U\subseteq (\complex^n)^{\otimes m}$ of dimension $(1-o(1))(1-\varepsilon) n^m$ with $E(\U)\geq {\varepsilon m^{-m}}$. In the special case that $\eps={m d\choose d,d,\dots,d}^{-1}$ for some $d \in \natural$, this bound can be strengthened to $E(\U) \geq {\eps}$.
 \end{theorem}
 \begin{proof}
     Choose $d$ such that
     \begin{equation}\label{eq:kdepsilon}
     {md\choose d,d,\dots,d}^{-1}\leq \varepsilon<{m(d-1)\choose (d-1),(d-1),\dots,(d-1)}^{-1}.
     \end{equation}
Multiplying by ${m d\choose d,d,\dots,d}$  gives
$$
1\leq {m d\choose d,d,\dots,d}\varepsilon<
\frac{(m d)(m d-1)\cdots (m d-m+1)}{d^m}\leq m^m.
$$
From Proposition~\ref{prop:D1bound} and (\ref{eq:kdepsilon}) it follows that
$$
E(\U)\geq {m d\choose d,d,\dots,d}^{-1}\geq {\varepsilon m^{-m}}.
$$
In the special case ${m d\choose d,d,\dots,d}\varepsilon=1$ this bound can be strengthened to $E(\U)\geq \eps$, as claimed. The dimension of $\U$ is at least
\ba
n^m- n^m{m d\choose d,d,\dots,d}^{-1}\left(1+\frac{m e}{\sqrt[d]{n}}\right)^{m d}&>n^m - \varepsilon n^m \left(1+\frac{m e}{\sqrt[d]{n}}\right)^{m d}\\
& \geq (1-o(1))(1-\eps)n^m.
\ea
This completes the proof.
 \end{proof}
The following theorem gives a general tradeoff between $\dim(\U)$ and $E(\U)$. Here, $\ln(\cdot)$ denotes the logarithm base $e$.
\begin{theorem}\label{thm:maxnorank1general}
Let $0<\alpha<m$ be arbitrary but fixed. There exists a
sequence of subspaces $\U_n\subseteq (\complex^n)^{\otimes m}$, $n\geq 1$ of dimension at least $n^m-n^{f_m(\alpha)}$ with $E(\U_n) \geq n^{-2\alpha}$, where
$$f_m(\alpha)=m+\frac{2\alpha\ln(m^{-1}+e m^{- m/2\alpha})}{\ln m}.
$$
\end{theorem}
\begin{proof}
Note that
$$
\lim_{d\to\infty}\frac{\ln{md\choose d,d,\dots,d}}{d}=m\ln m.
$$
Define $d=d(n)=\lceil \beta\ln(n)\rceil$, where $\beta=2\alpha/(m\ln m)$.
Then we have
$$
\lim_{n\to\infty}\frac{\ln{md(n)\choose d(n),d(n),\dots,d(n)}}{\ln n}= \lim_{n\to \infty}\frac{d(n)}{\ln n}\cdot \frac{\ln {md(n)\choose d(n),d(n),\dots,d(n)}}{d(n)}\geq  \beta m\ln m.
$$
Let $\U_n$ be equal to $\U$ as in Proposition~\ref{prop:D1bound}.
Then we have
 $$
\liminf_{n\to \infty}\frac{\ln E(\U_n)}{\ln(n)}\geq \liminf_{n\to \infty} \frac{\ln {m d(n)\choose d(n),d(n),\dots,d(n)}^{-1}}{\ln n}=-\beta m \ln m=-2 \alpha,
$$
which proves that $E(\U_n) \geq n^{-2\alpha}$. Let $c(n)=n^m-\dim(\U_n)$ be the codimension of $\U_n$. From
$$
c(n)\leq n^m{md(n)\choose d(n),d(n),\dots,d(n)}^{-1}\Big(1+\frac{me}{\sqrt[d(n)]{n}}\Big)^{md(n)}
$$
it follows that
\begin{multline*}
\limsup_{n\to\infty}\frac{\ln c(n)}{\ln(n)}\leq m-\beta m\ln m +\beta m\ln(1+me^{1-1/\beta})=m+\beta m\ln(m^{-1}+e^{1-1/\beta})=\\
=m+\frac{2\alpha\ln(m^{-1}+e\cdot e^{- m\ln(m)/2\alpha})}{\ln m}=
m+\frac{2\alpha\ln(m^{-1}+e m^{- m/2\alpha})}{\ln m}.
\end{multline*}
This completes the proof.
\end{proof}

\begin{remark}\label{rmk:rank2}
We remark that our subspace $\U_{a,d_1,d_2} := S^{d_1+d_2}(\complex^a)^{\perp} \subseteq S^{d_1}(\complex^a) \otimes S^{d_2}(\complex^a)$ always contains elements of Schmidt rank 2. Assume without loss of generality $d_1 \leq d_2$, and define unit vectors
\[
e_1 := \ket{1}^{d_1} \in S^{d_1} (\complex^a) \qquad f_1 := \ket{2}^{d_2} \in S^{d_2}(\complex^a)
\]
\[
e_2 := \ket{2}^{d_1} \in S^{d_1}(\complex^a),\qquad f_2 := \sqrt{\binom{d_2}{d_1}}\;\ket{1}^{d_1}\ket{2}^{d_2-d_1} \in S^{d_2}(\complex^a).
\]
Then $\Pi_{a,d_1,d_2}(e_1 \otimes f_1)$ and ${\Pi_{a,d_1,d_2}(e_2 \otimes f_2)}$ are nonzero vectors in $S^{d_1+d_2}(\complex^a)$ that span a 1-dimensional space, so there exists a non-zero scalar $\alpha$ for which $e_1 \otimes f_1+\alpha e_2 \otimes f_2 \in \ker(\Pi_{a,d_1,d_2})=\U$. This is a Schmidt rank 2 vector in $\U$. A straightforward calculation shows that one can take $\alpha=-\binom{d_2}{d_1}^{-1/2}$.
\end{remark}

\section{Further applications}\label{sec:applications}

In this section we discuss further applications of our construction to produce entangled mixed states that remain entangled after perturbation, and entanglement witnesses with many, highly negative eigenvalues.

In the following, $\norm{\cdot}_1$ denotes the trace norm and $\norm{\cdot}_{\infty}$ denotes the spectral norm.
\begin{theorem}[Robustly entangled mixed states]
Let $\H=(\complex^n)^{\otimes m}$.
\begin{enumerate}
\item There is an explicit mixed state $\rho$ on $\H$ of rank $n^m-m(n-1)-1$ such that for every Hermitian operator $H$ with trace norm $\norm{H}_1 \leq m^{-(n-1)m/2}$, the state $e^{iH} \rho e^{-iH}$ is entangled.
\item For any $0 < \eps < 1$, there is an explicit mixed state $\rho$ on $\H$ of rank $(1-o(1))(1-\eps)n^m$ such that $e^{iH} \rho e^{-iH}$ is entangled for every Hermitian operator $H$ with trace norm $\norm{H}_1 \leq \sqrt{\eps m^{-m}}$. In the particular case when $\eps=\binom{md}{d,\dots,d}^{-1}$ for some positive integer $d$, this condition can be weakened to $\norm{H}_1 \leq \sqrt{\eps}$.
\end{enumerate}
\end{theorem}
\begin{proof}
This follows from~\cite[Theorem 2]{zhu2024quantifying} by taking $\rho$ to be a state supported on one of the subspaces specified by Theorem~\ref{thm:main}.
\end{proof}

\begin{theorem}[Entanglement witnesses with many, highly negative eigenvalues]\label{thm:witness_construction}
Let $\H=(\complex^n)^{\otimes m}$.
\begin{enumerate}
\item There is an explicit entanglement witness $H$ with $\norm{H}_{\infty}=1$ having $n^m-m(n-1)-1$ negative eigenvalues, each of magnitude at least $m^{-(n-1)m}$.
\item For any $0 < \eps < 1$ there is an explicit entanglement witness $H$ with $\norm{H}_{\infty}=1$ having $(1-o(1))(1-\eps)n^m$ negative eigenvalues, each of magnitude at least ${\varepsilon m^{-m}}$. In the particular case when $\eps=\binom{md}{d,\dots,d}^{-1}$ for some positive integer $d$, this bound on the magnitude can be improved to $\eps$.
\end{enumerate}
\end{theorem}
\begin{proof}
This follows from e.g.~\cite[Proposition 5.4]{DJL24} by taking $H=\I-\mu \Pi_{\U}$, where $\Pi_{\U}$ is the projection onto one of the subspaces specified in Theorem~\ref{thm:main} and $\mu=(1-E(\U))^{-1}$.
\end{proof}

\section*{Acknowledgments}

We thank Felix Leditzky and Debbie Leung for helpful discussions. H.D. was supported by the National Science Foundation under Grant No. DMS-2147769. B.L. acknowledges that part of this work was funded by the National Science Foundation under Award No. DMS-2202782.

\bibliographystyle{alpha}
\bibliography{references}

\end{document}